\documentclass[english,american,aps,pre,twocolumn]{revtex4-1}
\usepackage[T1]{fontenc}
\usepackage[latin9]{inputenc}
\usepackage{verbatim}
\usepackage{textcomp}
\usepackage{amstext}
\usepackage{amssymb}
\usepackage{graphicx}

\makeatletter

\@ifundefined{textcolor}{}
{%
 \definecolor{BLACK}{gray}{0}
 \definecolor{WHITE}{gray}{1}
 \definecolor{RED}{rgb}{1,0,0}
 \definecolor{GREEN}{rgb}{0,1,0}
 \definecolor{BLUE}{rgb}{0,0,1}
 \definecolor{CYAN}{cmyk}{1,0,0,0}
 \definecolor{MAGENTA}{cmyk}{0,1,0,0}
 \definecolor{YELLOW}{cmyk}{0,0,1,0}
}

\usepackage{graphics,psfrag}
\usepackage{graphicx,psfrag}

\makeatother

\usepackage{babel}
\begin{document}

\title{Fluids confined in wedges and by edges:\\
 From cluster integrals to thermodynamic properties referred to different
regions}

\author{Ignacio Urrutia$^{\dag}$}

\email{iurrutia@cnea.gov.ar}

\affiliation{$^{\dag}$Departamento de Física de la Materia Condensada, Centro
Atómico Constituyentes, CNEA, Av.Gral.~Paz 1499, 1650 Pcia.~de Buenos
Aires, Argentina}

\affiliation{CONICET}
\begin{abstract}
Recently, new insights in the relation between the geometry of the
vessel that confines a fluid and its thermodynamic properties were
traced through the study of cluster integrals for inhomogeneous fluids.
In this work I analyze the thermodynamic properties of fluids confined
in wedges or by edges, emphasizing on the question of the region to
which these properties refer. In this context, the relations between
the line-thermodynamic properties referred to different regions are
derived as analytic functions of the dihedral angle $\alpha$, for
$0<\alpha<2\pi$, which enables a unified approach to both edges and
wedges. As a simple application of these results, I analyze the properties
of the confined gas in the low-density regime. Finally, using recent
analytic results for the second cluster integral of the confined hard
sphere fluid, the low density behavior of the line thermodynamic properties
is analytically studied up to order two in the density for $0<\alpha<2\pi$
and by adopting different reference regions.
\end{abstract}
\maketitle

\section{Introduction\label{sec:Intro}}

The interest on confined inhomogeneous fluids covers a large length
of scales of the particles size which starts at the simplest one-atom
per molecule (e.g. the noble gases) and goes up to proteins, polymers
(including DNA molecules), and large colloids.\cite{Henderson_2006,Almenar_2011,Karl_2011,Lutsko_2012_b,Statt_2012}
The thermodynamic properties of these systems are influenced by the
geometry of the vessel or substrate that constrains the spatial region
where the molecules of the system are enabled to move. Important efforts
are continuously devoted to reach a detailed description of the response
of fluids to some simple geometrical constraints like the confinement
in pores with slit, cylindrical and spherical shapes, as well as the
case of fluids in contact with planar and curved walls. 

This work focuses on fluids confined by open dihedrons built by two
planar faces that meet in an edge. Previous studies were dedicated
to analyze the adsorption of liquid-vapor coexisting phases on edges
and wedges,\cite{Rejmer_1999,Henderson_2002,Henderson_2004,Henderson_2004_b}
and also, to the adsorption on corrugated surfaces.\cite{Schneemilch_2003,Bryk_2003b,Schoen_2002}
The main characteristic of the thermodynamics of fluids confined by
edges and wedges is the existence of line tensions. This property
also characterizes systems with two coexisting phases adsorbed on
planar substrates (sessile drops) and systems with three coexisting
phases that meet on a common line.\cite{Getta_1998,Schimmele_2007}

One of the particularities of the edge/wedge type of confinement is
that it produces non-trivial spatial inhomogeneities of the fluid.
As in the case of fluids confined by curved walls, it happens that
different points of view in the very beginning of the analysis produce
dissimilar properties.\cite{Urrutia_2014} Thus, it is relevant to
establish the basis that allow us to compare the thermodynamic properties
found by adopting these different points of view.

In this work, I analyze the statistical mechanics and thermodynamic
properties of a fluid confined in an edge/wedge on the basis of the
representation of its grand potential in powers of the activity.
In Sec. \ref{sec:ClusterinPolyhedra} different type of edge/wedge
confinements are discussed and the thermodynamics of the fluid composed
by spherical particles is revisited. There, I analyze the free energy
and related thermodynamic magnitudes of the confined fluid emphasizing
on the explicit choice of the reference region to which system properties
refer. Sec. \ref{sec:WedgeThermo} describes the functional dependence
of the cluster integrals with the measures of the edge/wedge spatial
region and the consequences that follow on system properties. There,
new relations between bulk- surface- and line-thermodynamic properties
for different reference regions, are shown. They take the form of
transformation laws and apply to any density. Also, the behavior of
low density gases is discussed. In Sec. \ref{sec:HSlowdens}, this
approach is utilized to derive analytic expressions for the thermodynamic
properties (pressure, surface tension, line tension, surface- and
linear- adsorptions) of the confined hard sphere fluid up to order
two in density. The consequences of adopting different reference regions
on the thermodynamics of this system are also discussed in this section.
The expressions obtained of line-tension and line-adsorption, show
the dependence with the opening dihedral angle. Final remarks are
presented in Sec. \ref{sec:Conclusions}.

\section{Detailed description of a fluid in an edge/wedge confinement\label{sec:ClusterinPolyhedra}}

Let us consider an open system of particles at constant temperature
$T$ and chemical potential $\mu$, which is confined by two planar
walls that intersect in an edge. For simplicity we only refer here
to spherical particles. The walls exert a hard potential $\phi\left(\mathbf{r}\right)$
that constrains the position of the center of each particle to a region
$\mathcal{A}$ with dihedral shape (throughout this work the open-dihedron
geometrical shape is referred to as dihedron) being $\alpha$ the
inner angle between faces (inner to $\mathcal{A}$). %
I analyze two different types of edge/wedge confinement.
\begin{figure}
\begin{centering}
\includegraphics[height=3.3cm]{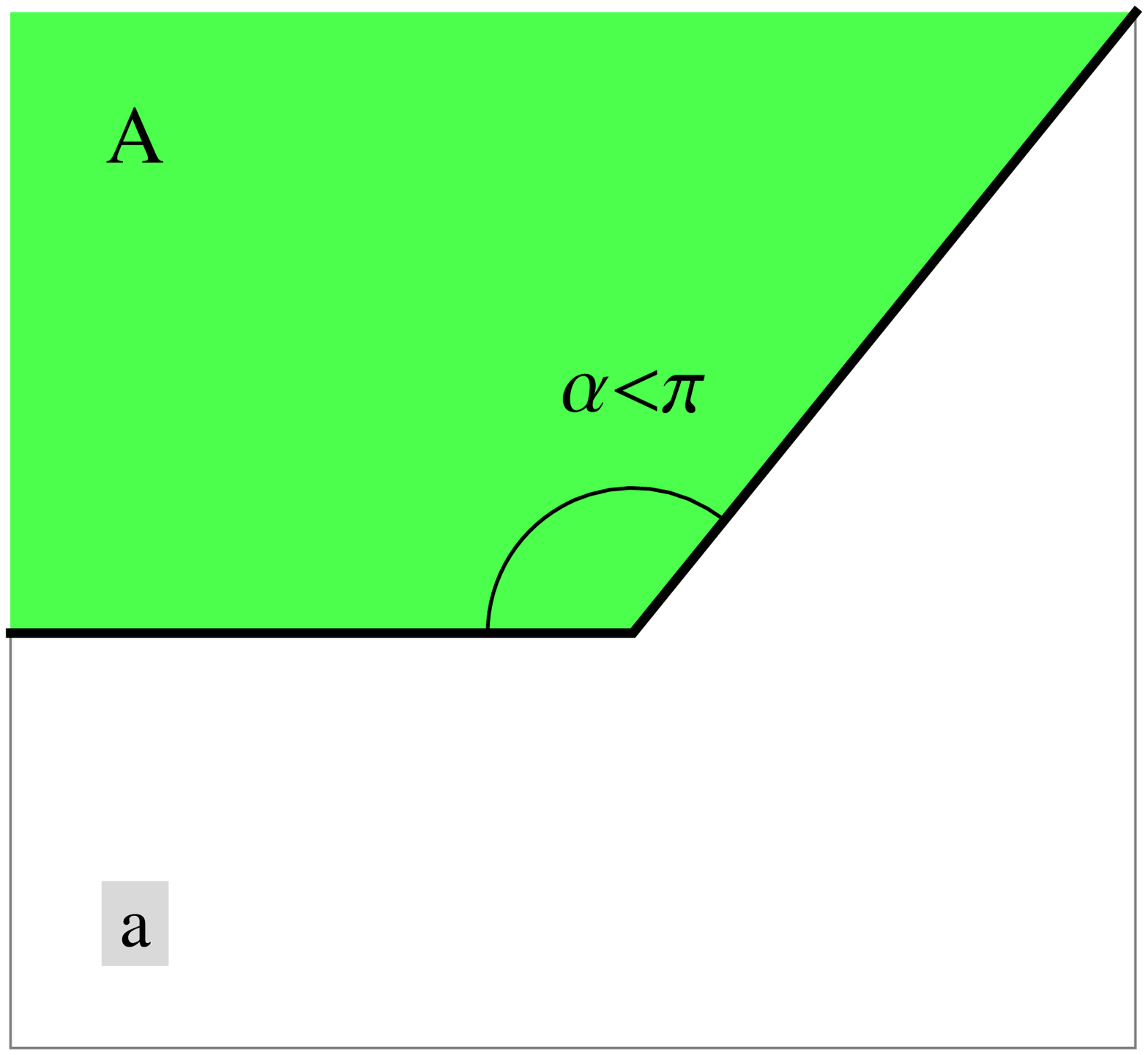}\includegraphics[height=3.3cm]{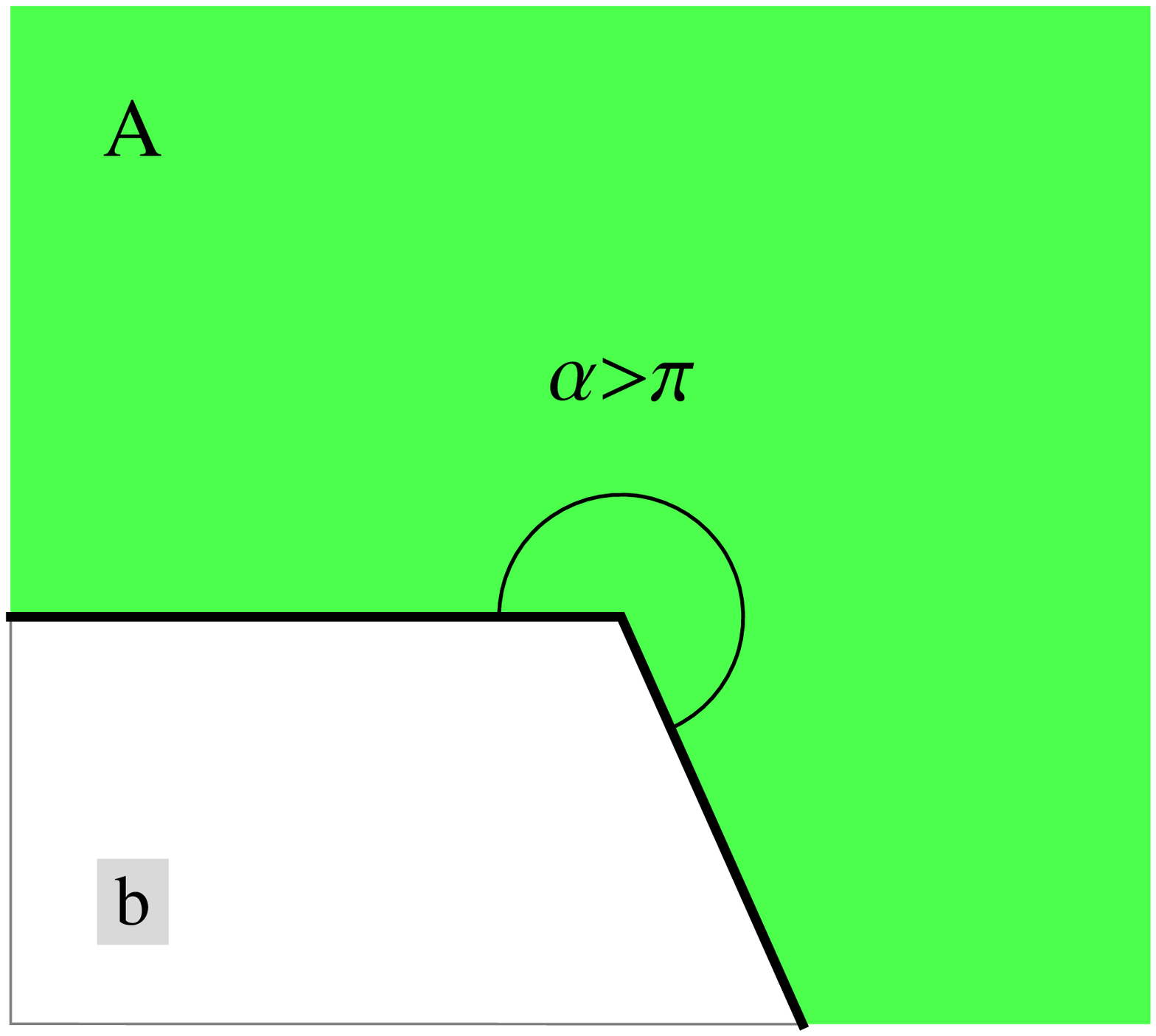}
\par\end{centering}

\protect\caption{Fluid confined by a hard-wall dihedron. In the region $\mathcal{A}$
in light-gray (green) particles are free to move while the region
in white is forbidden. Note that no matter the value of the opening
angle both light-gray (green) and white regions are straight-edge
dihedrons.\label{fig:Adihedron}}
\end{figure}
 Fig.\ref{fig:Adihedron} shows one of the edge/wedge confinement
considered. There, the edge/wedge available region $\mathcal{A}$
is defined by the Boltzmann factor $\exp\left[-\phi\left(\mathbf{r}\right)/kT\right]=\Theta\bigl(\left|\mathbf{r}-\mathcal{C}\right|\bigr)$
where \foreignlanguage{english}{$k$} is the Boltzmann's constant,
$\Theta\left(x\right)$ the Heaviside function {[}$\Theta\left(x\right)=1$
if $x>0$ and zero otherwise{]}, the dihedral region which is the
complement of $\mathcal{A}$ is $\mathcal{C}=\mathbb{R}^{3}\setminus\mathcal{A}$,
and $\left|\mathbf{r}-\mathcal{C}\right|$ is the shortest distance
between $\mathbf{r}$ and $\mathcal{C}$. Note that the faces of $\mathcal{A}$
meet on a straight line, thus, I call it a straight-edge dihedron.
\begin{figure}
\begin{centering}
\includegraphics[height=4cm]{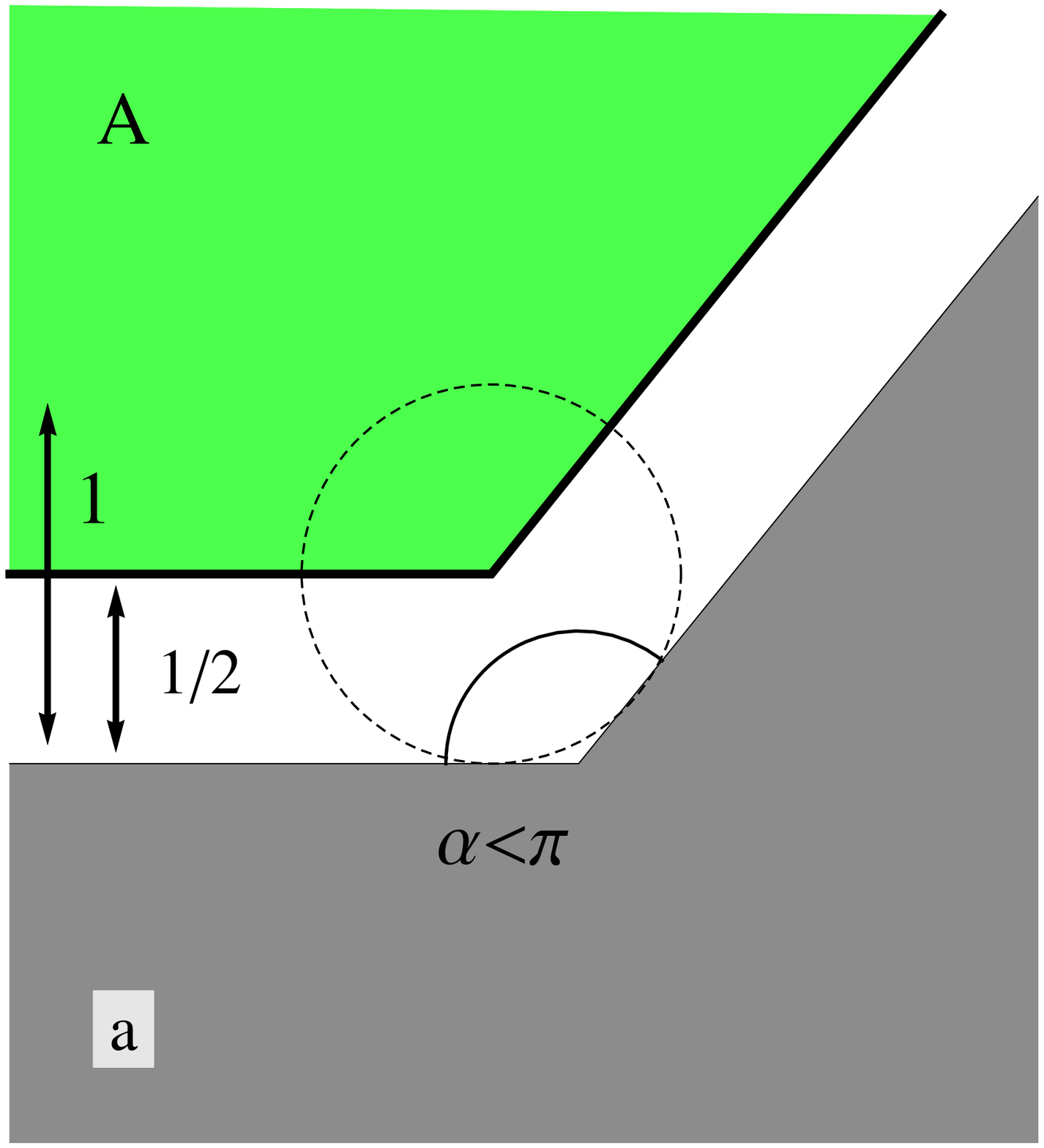}\includegraphics[height=4cm]{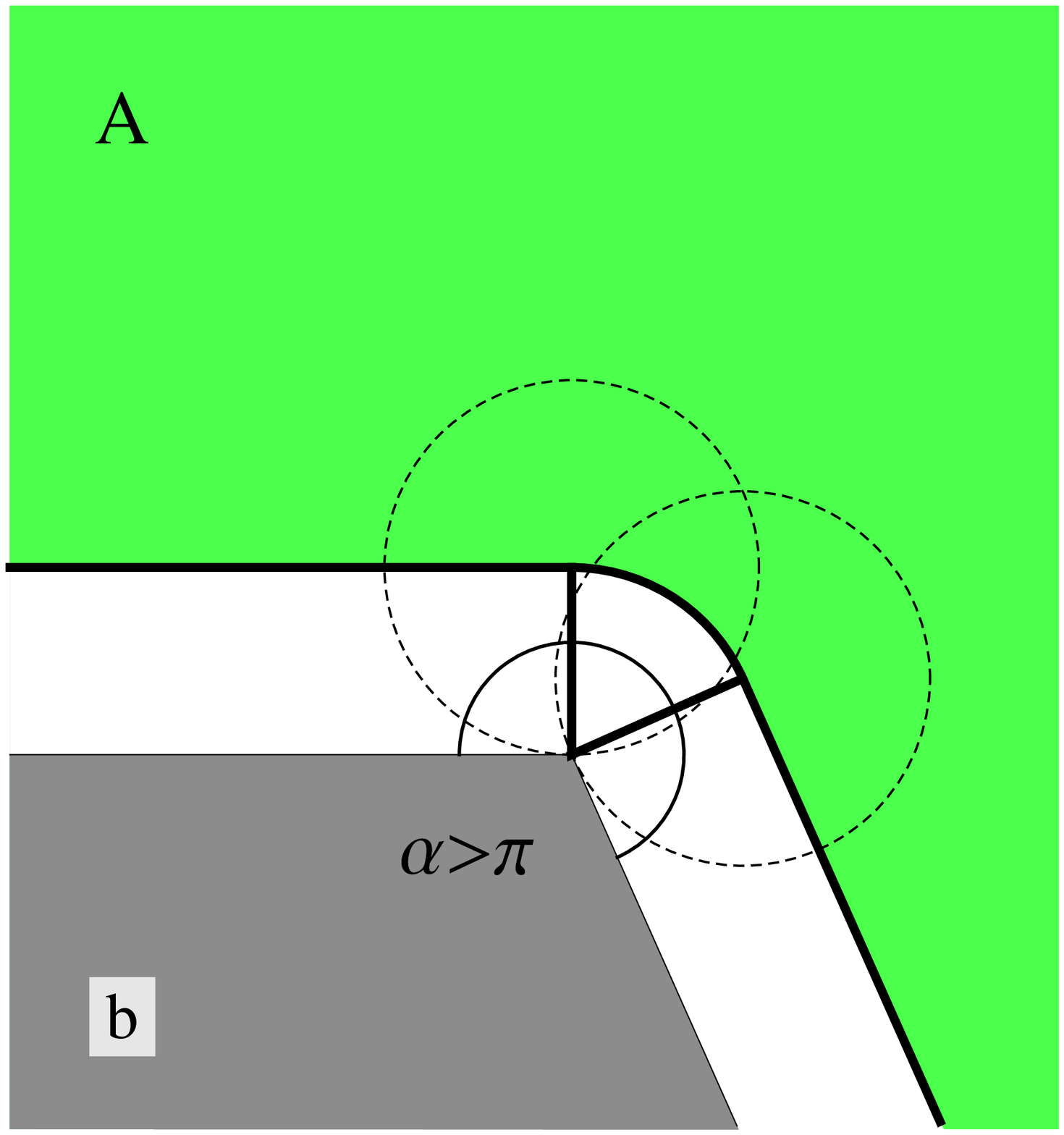}
\par\end{centering}

\protect\caption{Fluid confined by a hard wall dihedron that induce an excluded region.
Painted with darker gray is the hard wall, lighter gray (green) corresponds
to the fluid and the excluded region is in white. (a) shows
the case $\alpha<\pi$ while (b) corresponds to the case $\alpha>\pi$.
The wall-particle hard repulsion distance is $\sigma/2$ and dashed
circles represent particles at selected positions near the edge/wedge.
The arrows show characteristic lengths in $\sigma$ units.\label{fig:HWallEdge}}
\end{figure}
 On the other hand, one has the confinement defined by the Boltzmann
factor $\exp\left[-\phi\left(\mathbf{r}\right)/kT\right]=\Theta\left(\left|\mathbf{r}-\mathcal{C}\right|-\frac{\sigma}{2}\right)$
being $\mathcal{C}$ a solid dihedral region and $\sigma/2$ the minimum
distance between the center of a particle and the solid dihedron.
The latter case, particularly relevant for colloidal particles and
macromolecules, is drawn in Fig. \ref{fig:HWallEdge} where the forbidden
region between $\mathcal{A}$ and $\mathcal{C}$ is also indicated.
Note that for $0<\alpha<\pi$ the region $\mathcal{A}$ is a straight-edge
dihedron (shown in Fig. \ref{fig:HWallEdge}a). On the contrary, for
$\pi<\alpha<2\pi$ (shown in Fig. \ref{fig:HWallEdge}b) $\mathcal{A}$
is a rounded-edge dihedron (it has a curved end-of-fluid surface).
This kind of rounded edge confinement is produced by the external
hard potential being the inter-particles potential arbitrary. In summary,
Figs. \ref{fig:Adihedron}a, \ref{fig:Adihedron}b and \ref{fig:HWallEdge}a
correspond to a straight-edge confinement while Fig. \ref{fig:HWallEdge}b
corresponds to a rounded-edge.

Before analysing the thermodynamic properties of this confined fluid
it is necessary to adopt a region $\mathcal{B}$ as the reference
region (RR).\cite{Urrutia_2014b,Urrutia_2014} Note that $\mathcal{B}$
may coincide or not with $\mathcal{A}$. I wish to underline that
in the study of confined fluids is crucial to clearly establish the
adopted RR which fix the \emph{position and shape} of its boundary
the surface of tension. This question is as important as to establish
the system of reference in the study of a mechanical system. Hence,
I adopt as RR the region $\mathcal{B}$ which nearly follows the shape
of $\mathcal{A}$ and has measures $\mathbf{M}_{\mathcal{B}}=\left(V,A,L\right)$
(being $V$, $A$ and $L$ the volume, surface area and length of
the edge of $\mathcal{B}$, respectively). A detailed analysis of
different prescriptions for $\mathcal{B}$ is presented in Sec.\ref{sec:WedgeThermo}.
The grand potential of the confined fluid, relative to $\mathcal{B}$,
can be written as
\begin{equation}
\Omega=-PV+\gamma A+\mathcal{T}L\:,\label{eq:Omega}
\end{equation}
where $P$ is the pressure of the fluid, $\gamma$ is the wall/fluid
surface tension (or surface free-energy), and $\mathcal{T}$ the wall/fluid
line tension (or line free-energy). The mean number of particles in
the confined system is
\begin{eqnarray}
N & = & -\frac{\partial\Omega}{\partial\mu}\,,\label{eq:NyOmega}\\
 & = & \rho V+\Gamma_{A}A+\Gamma_{L}L\,,\label{eq:N}
\end{eqnarray}
where $\rho$ is the mean number density, $\Gamma_{A}$ is the excess
adsorption per unit area, and $\Gamma_{L}$ is the excess adsorption
per unit length. Naturally, this kind of linear decomposition also
applies to other magnitudes, such as the entropy $S=-\frac{\partial\Omega}{\partial T}$,
the energy $U=\Omega+TS+\mu N$, and higher order derivatives like
$\sigma_{_{N}}^{\,2}\equiv\left\langle N^{2}\right\rangle -N^{2}=kT\frac{\partial N}{\partial\mu}$
which describe fluctuations.

From Eq. (\ref{eq:Omega}) is clear that once $\mathcal{A}$ is fixed,
$\Omega$ becomes independent of the adopted RR. On the contrary,
since measures are relative to $\mathcal{B}$, some of the magnitudes
$(P,\gamma,\mathcal{T})$ depend on $\mathcal{B}$. Naturally, the
same argument shows that $N$ is independent of the adopted RR although
some of the magnitudes $(\rho,\Gamma_{A},\Gamma_{L})$ may depend
on $\mathcal{B}$, and so on. In summary, even when we have an idea
of the meaning of the magnitudes $(P,\gamma,\mathcal{T},\rho,\Gamma_{A},\Gamma_{L})$
that allow us to give a name to each one, they were not appropriately
defined yet. Indeed, they do not describe the pure properties of the
confined fluid (as it may be suggested by the adopted names for these
magnitudes), but they describe the properties of the fluid with regard
to a given RR in a sense that will be clarified below. Some general
aspects of the former discussion follow the analysis of macroscopic
systems with coexisting phases done in Ref. \cite{Schimmele_2007}.

\section{Cluster integrals and thermodynamics\label{sec:WedgeThermo}}

Let us consider a system of particles interacting through a pair potential
$\psi\left(r\right)$ with finite range. The center of these particles
is confined by a hard external potential to an edge/wedge region $\mathcal{A}$.
It was recently shown that the cluster integrals of the system take
the form\cite{Urrutia_2013_prep,Hill1956}
\begin{equation}
\tau_{i}=i!b_{i}V-i!a_{i}A+i!c_{i}L\:.\label{eq:Taui}
\end{equation}
Here, the $i$-th cluster integral $\tau_{i}$ is linear on the extensive
measures $\mathbf{M}_{\mathcal{A}}=\left(V,A,L\right)$ that geometrically
characterize the region $\mathcal{A}$ (its volume $V$, surface area
$A$, and the length of its edge $L$). We say that $\mathbf{M}_{\mathcal{A}}$
are the measures of the system relative to $\mathcal{A}$. Besides,
cluster integral also depends on the opening dihedral angle between
faces, $\alpha$. The volume coefficients $b_{i}$ are the well known
Mayer's cluster integrals for homogeneous systems and the area coefficients
$a_{i}$ were introduced to describe a fluid adsorbed on an infinite
wall.\cite{Bellemans_1962,Yang_2013} $b_{i}$ and $a_{i}$ with $i>1$
depend on $\psi\left(r\right)$ but are independent of $\alpha$,
being $\tau_{1}=b_{1}V$ with $b_{1}=1$ and $V=Z_{1}$, the configuration
integral of one particle. Eq. (\ref{eq:Taui}) was originally derived
for the case of a straight-edge dihedral region\cite{Urrutia_2013_prep}
$\mathcal{A}$, i.e. the cases described in Figs. \ref{fig:Adihedron}a,
\ref{fig:Adihedron}b and \ref{fig:HWallEdge}a, and was latter extended
to the rounded-edge dihedron shown in Fig.\ref{fig:HWallEdge}b.\cite{Urrutia_2014b} 

It is well known that in the low density regime or gaseous phase the
properties of the confined fluid can be rigorously written as power
series in the activity $z=\Lambda^{-3}\exp\!\left(\beta\mu\right)$
(here $\beta=1/kT$ is the inverse temperature and $\Lambda$ the
de Broglie\textasciiacute s thermal length).\cite{Hill1956} The Mayer
series of the grand potential for the confined fluid is given by 
\begin{equation}
\Omega=-\beta^{-1}\sum_{i\geq1}\frac{\tau_{i}}{i!}z^{i}\,,\label{eq:Omegaz}
\end{equation}
being its mean number of particles

\begin{equation}
N=\sum_{i\geq1}i\frac{\tau_{i}}{i!}z^{i}\,.\label{eq:Nz}
\end{equation}
By replacing Eq. (\ref{eq:Taui}) in Eqs. (\ref{eq:Omegaz}) and (\ref{eq:Nz}),
one obtains
\begin{equation}
\beta\Omega=-\Bigl(\sum_{i\geq1}b_{i}z^{i}\Bigr)V+\Bigl(\sum_{i\geq1}a_{i}z^{i}\Bigr)A-\Bigl(\sum_{i\geq1}c_{i}z^{i}\Bigr)L\:,\label{eq:Omegaz2}
\end{equation}
\begin{equation}
N=-\Bigl(\sum_{i\geq1}ib_{i}z^{i}\Bigr)V-\Bigl(\sum_{i\geq1}ia_{i}z^{i}\Bigr)A+\Bigl(\sum_{i\geq1}ic_{i}z^{i}\Bigr)L\:.\label{eq:Nz2}
\end{equation}
Consistently, a similar transformation applies to other thermodynamic
magnitudes {[}$S$, $U$, $\sigma_{_{N}}^{\,2}$, see Eq. (\ref{eq:N}){]}.
Eqs. (\ref{eq:Omega}) and (\ref{eq:Omegaz2}) have a similar dependence
with the measures. Nevertheless, Eq. (\ref{eq:Omega}) is in terms
of measures $\mathbf{M}_{\mathcal{B}}$ that correspond to the choice
$\mathcal{B}$ as RR, while Eq. (\ref{eq:Omegaz2}) is in terms of
measures regarding to $\mathcal{A}$ as RR. Of course, the same applies
to Eqs. (\ref{eq:N}) and (\ref{eq:Nz2}). Eqs. (\ref{eq:Omega}-\ref{eq:N})
show the advantages in deriving $\tau_{i}$ as a linear function of
the measures of $\mathcal{B}$. If it is possible, one find the $z$
power series for the intensive thermodynamic properties $P,\gamma,\mathcal{T}$
and densities $\rho,\Gamma_{A},\Gamma_{L}$, referred to $\mathcal{B}$.

Eq. (\ref{eq:Taui}) transpires lineal algebra concepts. It can be
seen as a vector in an abstract space with basis of coordinates $\left(V,A,L\right)$
and components $\left(i!b_{i},-i!a_{i},i!c_{i}\right)$ but also
as the inner product $\left(i!b_{i},-i!a_{i},i!c_{i}\right)\cdot\mathbf{M}_{\mathcal{A}}$
between row and column vectors that live in dual spaces. These analogies
flow to Eqs. (\ref{eq:Omegaz2}, \ref{eq:Nz2}) and will be further
investigated in the following Secs. \ref{sub:Case-I} and \ref{sub:Case-II}
for different ranges of $\alpha$.

The confinement of the systems drawn in Fig. \ref{fig:Adihedron}
is purely characterized by the region $\mathcal{A}$ where the density
distribution is non-null. This density-based choice of $\mathcal{B}$
will be labeled with a d subindex (d-RR). For the confinement shown
in Figs. \ref{fig:Adihedron}a and \ref{fig:Adihedron}b the unique
simple choice for RR is $\mathcal{A}$ itself, this prevents to analyze
them from the point of view of the freedom to choose the RR.

On the other hand, even when the systems shown in Fig. \ref{fig:HWallEdge}
can also be analyzed under the same density-based $\mathcal{B}$,
other RR could be adopted. To analyze this problem and the relation
between the thermodynamic properties obtained under different choices
of $\mathcal{B}$, we study the edges/wedges with angles $0<\alpha<\pi$
and $\pi<\alpha<2\pi$ separately. It is interesting to note that
when the d-RR is adopted the system depicted in Fig. \ref{fig:Adihedron}a
and the system drawn in Fig. \ref{fig:HWallEdge}a are identical,
and thus, their properties are identical too.

\subsection{Case I ($0<\alpha<\pi$)\label{sub:Case-I}}

For the case of the wedge confinement drawn in Fig. \ref{fig:HWallEdge}a,
I study two different choices for the RR that are the most natural
to be adopted. Under the density-based RR that identifies $\mathcal{B}$
with $\mathcal{A}$, the measures are $\mathbf{M}_{\textrm{d}}=\left(V_{\textrm{d}},A_{\textrm{d}},L\right)$,
being the $i$-th cluster integral
\begin{equation}
\tau_{i}/i!=b_{i}V_{\textrm{d}}-a_{i}A_{\textrm{d}}+c_{i}\!\left(\beta\right)L=\mathbf{b}_{\textrm{d}}\cdot\mathbf{M}_{\textrm{d}}\:,\label{eq:TauiMd}
\end{equation}
where $\mathbf{b}_{\textrm{d}}=\left(b_{i},-a_{i},c_{i}\right)$ is
the vector of coefficients. The second simple choice for $\mathcal{B}$
is the empty-region (e-RR), i.e. $\mathcal{B}$ is taken as $\mathcal{A}$
joined with the white region in Fig. \ref{fig:HWallEdge}a and the
measures are $\mathbf{M}_{\textrm{e}}=\left(V_{\textrm{e}},A_{\textrm{e}},L\right)$.
In what follows we will use $\sigma$ (see Fig. \ref{fig:HWallEdge})
as the unit length. From geometrical considerations it is possible
to obtain the linear relation between both sets of measures: $V_{\textrm{d}}=V_{\textrm{e}}-\frac{1}{2}A_{\textrm{e}}+\frac{1}{4}\cot\!\frac{\alpha}{2}\, L$
for $0<\alpha<\pi$, $A_{\textrm{d}}=A_{\textrm{e}}-\csc\!\alpha\, L$
for $0<\alpha<\frac{\pi}{2}$, and $A_{\textrm{d}}=A_{\textrm{e}}-\cot\!\frac{\alpha}{2}\, L$
for $\frac{\pi}{2}<\alpha<\pi$. Thus, under the e-RR choice
\begin{equation}
\tau_{i}/i!=\tilde{b}_{i}V_{\textrm{e}}-\tilde{a}_{i}A_{\textrm{e}}+\tilde{c}_{i}\!\left(\alpha\right)L=\mathbf{b}_{\textrm{e}}\cdot\mathbf{M}_{\textrm{e}}\:.\label{eq:TauiMe}
\end{equation}
We introduce the matrix $Y$ that transforms between both sets of
measures
\begin{equation}
\mathbf{M}_{\textrm{d}}=Y\cdot\mathbf{M}_{\textrm{e}}\,\textrm{ and }\,\mathbf{M}_{\textrm{e}}=Y^{-1}\cdot\mathbf{M}_{\textrm{d}}\:.\label{eq:MdMe}
\end{equation}
Its expression follows from the relations above Eq. (\ref{eq:TauiMe})
\begin{equation}
Y=\left(\begin{array}{ccc}
1 & -\frac{1}{2} & \frac{1}{4}\cot\!\frac{\alpha}{2}\\
0 & 1 & -y(\alpha)\\
0 & 0 & 1
\end{array}\right)\:,\label{eq:YmatMde}
\end{equation}
with $y(\alpha)=\csc\!\alpha$ if $0<\alpha<\frac{\pi}{2}$ and $y(\alpha)=\cot\!\frac{\alpha}{2}$
if $\frac{\pi}{2}<\alpha<\pi$ (note that $y(\alpha)$ is a continuous
non-derivable function at $\alpha=\frac{\pi}{2}$). Given that $\Omega$
remains unmodified no matter which RR is adopted, one finds the linear
relation between the unknown coefficients $\mathbf{b}_{\textrm{e}}=\left(\tilde{b}_{i},-\tilde{a}_{i},\tilde{c}_{i}\right)$
and the known $\mathbf{b}_{\textrm{d}}$, through the $Y$ matrix
\begin{eqnarray}
\mathbf{b}_{\textrm{e}}\cdot\mathbf{M}_{\textrm{e}} & = & \mathbf{b}_{\textrm{d}}\cdot Y\cdot Y^{-1}\cdot\mathbf{M}_{\textrm{d}}\:,\nonumber \\
\mathbf{b}_{\textrm{e}} & = & \mathbf{b}_{\textrm{d}}\cdot Y\:.\label{eq:bMed}
\end{eqnarray}
Besides, through Eqs. (\ref{eq:Omega}-\ref{eq:N}) $Y$ also transforms
the thermodynamic properties 
\begin{equation}
\left(-P,\gamma,\mathcal{T}\right)_{\textrm{e}}=\left(-P,\gamma,\mathcal{T}\right)_{\textrm{d}}\cdot Y\:,\label{eq:intensY}
\end{equation}
\begin{equation}
\left(\rho,\Gamma_{A},\Gamma_{L}\right)_{\textrm{e}}=\left(\rho,\Gamma_{A},\Gamma_{L}\right)_{\textrm{d}}\cdot Y\:,\label{eq:adsY}
\end{equation}
where the label outside brackets shows the adopted reference region.
These relations should be valid even when the series expansion in
powers of $z$ does not apply, and thus, they should apply to any
density. Eqs. (\ref{eq:intensY}, \ref{eq:adsY}), with $Y$ taken
from Eq. (\ref{eq:YmatMde}), are one of the main results of the current
work. They show the transformation law between the intensive properties
of the confined system when different RRs are adopted.

Now, we turn our attention to Eqs. (\ref{eq:Omega}, \ref{eq:Omegaz2})
and (\ref{eq:intensY}). They show that $P_{\textrm{e}}=P_{\textrm{d}}=P$
being $P$ the pressure of the \emph{bulk} fluid at the same $T$
and $\mu$. Furthermore, one obtains 
\begin{equation}
\gamma_{\textrm{e}}=\gamma_{\textrm{d}}+P/2\:.\label{eq:gamed}
\end{equation}
Eq. (\ref{eq:gamed}) found here for a wedge confinement is a known
relation for fluids adsorbed on both planar and curved walls.\cite{Yang_2013,Urrutia_2014}
The $z$ power series representation of $\gamma_{\textrm{e}}$ and
$\gamma_{\textrm{d}}$ shows that they are the surface tension of
the fluid in contact with an infinite planar wall (each one for a
different RR). The line tension transforms as
\begin{equation}
\mathcal{T}_{\textrm{e}}\!=\mathcal{T}_{\textrm{d}}\!-\gamma_{\textrm{d}}\, y(\alpha)-\frac{P}{4}\cot\frac{\alpha}{2}\:,\label{eq:linetened}
\end{equation}
To the best of my knowledge it is the first time that Eq. (\ref{eq:linetened}),
which applies to any density, is derived. Turning to Eqs. (\ref{eq:N},
\ref{eq:Nz2}) and (\ref{eq:adsY}), they imply that $\rho_{\textrm{e}}=\rho_{\textrm{d}}=\rho$
with $\rho$ the number density of the \emph{bulk} fluid (at the same
$T$ and $\mu$). For the surface adsorption one finds,
\begin{equation}
\left(\Gamma_{A}\right)_{\textrm{e}}=\left(\Gamma_{A}\right)_{\textrm{d}}-\rho/2\:,\label{eq:adsAed}
\end{equation}
which is known to be an exact relation for planar walls. Again, based
on the $z$ power series one finds that both $\left(\Gamma_{A}\right)_{\textrm{e}}$
and $\left(\Gamma_{A}\right)_{\textrm{d}}$ are the adsorption of
the fluid on an infinite planar wall (each one for a different RR).
Furthermore, one obtains for the excess linear adsorption
\begin{equation}
\left(\Gamma_{L}\right)_{\textrm{e}}=\left(\Gamma_{L}\right)_{\textrm{d}}-\left(\Gamma_{A}\right)_{\textrm{d}}\, y(\alpha)+\frac{\rho}{4}\cot\!\frac{\alpha}{2}\:.\label{eq:adsLed}
\end{equation}
Again, this expression applies to any density and it was never published
before.

\subsection{Case II ($\pi<\alpha<2\pi$)\label{sub:Case-II}}

Now, focusing on the case shown in Fig. \ref{fig:HWallEdge}b, I
will analyze two different choices for the region $\mathcal{B}$,
which split in three different sets of measures that are the most
natural to adopt. The first choice is the d-RR, which corresponds
to identify $\mathcal{B}$ with $\mathcal{A}$. For this d-RR one
can consider two different criteria to define the measures depending
on whether $A$ is taken as the area of the planar part (\foreignlanguage{english}{$A_{p\textrm{d}}$})
of the surface $\partial\mathcal{B}$ or as its total area. Thus,
using the first criteria the measures are $\mathbf{M}_{\textrm{d}1}=\left(V_{\textrm{d}},A_{p\textrm{d}},L\right)$
and
\begin{equation}
\tau_{i}/i!=b_{i}V_{\textrm{d}}-a_{i}A_{p\textrm{d}}+c_{i}\!\left(\alpha\right)L=\mathbf{b}_{\textrm{d}1}.\mathbf{M}_{\textrm{d}1}\:,\label{eq:TauiMd1}
\end{equation}
with the vector of coefficients $\mathbf{b}_{\textrm{d}1}=\left(b_{i},-a_{i},c_{i}\right)$.
However, if one adopts the second criteria that assumes $A$ as the
total area of $\partial\mathcal{B}$ it is obtained
\begin{equation}
\tau_{i}/i!=\bar{b}_{i}V_{\textrm{d}}-\bar{a}_{i}A_{\textrm{d}}+\bar{c}_{i}\!\left(\alpha\right)L=\mathbf{b}_{\textrm{d}2}.\mathbf{M}_{\textrm{d}2}\:,\label{eq:TauiMd2}
\end{equation}
with $\mathbf{M}_{\textrm{d}2}=\left(V_{\textrm{d}},A_{\textrm{d}},L\right)$
and $A_{p\textrm{d}}=A_{\textrm{d}}-\left(\alpha-\pi\right)\frac{1}{2}L$.
The relationship between both sets of measures is
\begin{equation}
\mathbf{M}_{\textrm{d}1}=Y\cdot\mathbf{M}_{\textrm{d}2}\:,\:\:\mathbf{M}_{\textrm{d}2}=Y^{-1}\cdot\mathbf{M}_{\textrm{d}1}\:,\label{eq:Md1}
\end{equation}
while the vectors of coefficients relate through
\begin{equation}
\mathbf{b}_{\textrm{d}2}=\mathbf{b}_{\textrm{d}1}\cdot Y\:,\label{eq:bd1}
\end{equation}
with 
\begin{equation}
Y=\left(\begin{array}{ccc}
1 & 0 & 0\\
0 & 1 & -\frac{1}{2}\left(\alpha-\pi\right)\\
0 & 0 & 1
\end{array}\right)\:.\label{eq:YmatMd1d2}
\end{equation}
The relations between the equations of state $\left(-P,\gamma,\mathcal{T}\right)$
and also $\left(\rho,\Gamma_{A},\Gamma_{L}\right)$, in d1-RR and
d2-RR are given by Eqs. (\ref{eq:intensY}) and (\ref{eq:adsY}) with
the obvious change of labels and with $Y$ taken from Eq. (\ref{eq:YmatMd1d2}).
Therefore, one finds $P_{\textrm{d}2}=P_{\textrm{d}1}=P$ (with $P$
the bulk pressure), $\gamma_{\textrm{d}2}=\gamma_{\textrm{d}1}$ (which
are equal to the planar-wall surface tension $\gamma_{\textrm{d}}$
discussed for the case $\alpha<\pi$) and 
\begin{equation}
\mathcal{T}_{\textrm{d}2}=\mathcal{T}_{\textrm{d}1}-\gamma_{\textrm{d}}\left(\alpha-\pi\right)/2\:.\label{eq:linetend2d1}
\end{equation}
Besides, it is obtained $\rho_{\textrm{d}2}=\rho_{\textrm{d}1}=\rho$,
$\left(\Gamma_{A}\right)_{\textrm{d}2}=\left(\Gamma_{A}\right)_{\textrm{d}1}$
(which are equal to the planar-wall adsorption $\left(\Gamma_{A}\right)_{\textrm{d}}$
discussed for the case $\alpha<\pi$) and 
\begin{equation}
\left(\Gamma_{L}\right)_{\textrm{d}2}=\left(\Gamma_{L}\right)_{\textrm{d}1}-\left(\Gamma_{A}\right)_{\textrm{d}1}\left(\alpha-\pi\right)/2\:.\label{eq:AdsLd2d1}
\end{equation}
It seems that Eqs. (\ref{eq:linetend2d1}) and (\ref{eq:AdsLd2d1}),
that apply to any density, are novel results.

The other choice for $\mathcal{B}$ is the e-RR, that is, the join
of region $\mathcal{A}$ and the white region in Fig. \ref{fig:HWallEdge}b.
In this case, the measures are $\mathbf{M}_{\textrm{e}}=\left(V_{\textrm{e}},A_{\textrm{e}},L\right)$
and
\begin{equation}
\tau_{i}/i!=\tilde{b}_{i}V_{\textrm{e}}-\tilde{a}_{i}A_{\textrm{e}}+\tilde{c}_{i}\!\left(\alpha\right)L=\mathbf{b}_{\textrm{e}}.\mathbf{M}_{\textrm{e}}\:,\label{eq:TauiMe2}
\end{equation}
with $V_{\textrm{d}}=V_{\textrm{e}}-\frac{1}{2}A_{\textrm{e}}-\frac{1}{8}\left(\alpha-\pi\right)L$
and $A_{p\textrm{d}}=A_{\textrm{e}}$. Eqs. (\ref{eq:Md1}, \ref{eq:bd1})
describe the transformation between both, the measures and the coefficients,
they remain valid with the change of labels $\textrm{d}2\rightarrow\textrm{e}$
and for 
\begin{equation}
Y=\left(\begin{array}{ccc}
1 & -\frac{1}{2} & -\frac{1}{8}\left(\alpha-\pi\right)\\
0 & 1 & 0\\
0 & 0 & 1
\end{array}\right)\:.\label{eq:YmatMd1e}
\end{equation}
Yet non-surprising, following Eqs. (\ref{eq:intensY}) and (\ref{eq:adsY})
with the obvious change of labels and taking $Y$ from Eq. (\ref{eq:YmatMd1e})
one obtains $P_{\textrm{e}}=P$ and $\rho_{\textrm{e}}=\rho$. Furthermore,
both $\gamma_{\textrm{e}}$ and $\left(\Gamma_{A}\right)_{\textrm{e}}$
coincide with the planar-wall magnitudes found for the case $\alpha<\pi$,
and then Eqs. (\ref{eq:gamed}, \ref{eq:adsAed}) apply for the broad
range $0<\alpha<2\pi$. Finally, one obtains
\begin{equation}
\mathcal{T}_{\textrm{e}}=\mathcal{T}_{\textrm{d}1}+P\left(\alpha-\pi\right)/8\:,\label{eq:linetened1}
\end{equation}
\begin{equation}
\left(\Gamma_{L}\right)_{\textrm{e}}=\left(\Gamma_{L}\right)_{\textrm{d}1}-\rho\left(\alpha-\pi\right)/8\:.\label{eq:AdsLed1}
\end{equation}
Eqs. (\ref{eq:linetened1}) and (\ref{eq:AdsLed1}) were not published
earlier.

\subsection{Low density\label{sub:Low-density}}

In this brief digression the confined ideal gas and the low density
regime of the confined non-ideal gas are analyzed. I first concentrate
in the d-RR (d-RR for $0<\alpha<\pi$, d1- and d2-RR for $\pi<\alpha<2\pi$).
To obtain the properties of the confined ideal gas one truncates all
the series in Eqs. (\ref{eq:Nz}, \ref{eq:Omegaz2}, \ref{eq:Nz2})
at the first order in power of $z$. By adopting d-RR the volume $V_{\textrm{d}}$
is equal to $Z_{1}$ and the first cluster integral is $\tau_{1}=V_{\textrm{d}}$.
Thus, $\beta P=z$, $\rho=z$,
\begin{equation}
\gamma=\mathcal{T}=0\:,\:\textrm{ and }\:\Gamma_{A}=\Gamma_{L}=0\:.\label{eq:gamGam0}
\end{equation}
Therefore, under d-RR the \emph{confined ideal gas} is thoroughly
described by $\beta P=\rho$, i.e. the equation of state of the \emph{bulk
ideal gas}. Clearly, if we turn to e-RR the edge/wedge confined ideal
gas has non-null surface- and line- free energies. They can be evaluated
using Eq. (\ref{eq:gamGam0}) and the transformations discussed in
Secs. \ref{sub:Case-II} and \ref{sub:Case-I}. The conclusion is
that in order to obtain the simpler expressions for the thermodynamics
of the confined ideal gas, the d-RR is better than e-RR.

For the confined non-ideal gas under d-RR the first cluster integral
remains unmodified in comparison with the ideal gas. The second and
higher order $\tau_{i}$ could be calculated by direct integration.
Now, the series given in Eqs. (\ref{eq:Omegaz}, \ref{eq:Nz}, \ref{eq:Omegaz2},
\ref{eq:Nz2}) are truncated at order two in $z$, which gives surface
and linear thermodynamic properties proportional to $z^{2}$. Using
Eqs. (\ref{eq:NyOmega}, \ref{eq:Taui}) and trivial series manipulation
one obtains the power series for $z\left(\rho\right)$ and the series
representation of the thermodynamic properties in powers of $\rho$.
Up to order $\rho^{2}$ it is obtained: $\beta P=\rho-b_{2}\rho^{2}$
(i.e. the virial series for the bulk gas \cite{Hill1956}) and 
\begin{equation}
\beta\gamma=-\Gamma_{A}/2\:,\quad\beta\mathcal{T}=-\Gamma_{L}/2\:\label{eq:inhomID}
\end{equation}
{[}with $\Gamma_{A}=-2a_{2}\rho^{2}$ and $\Gamma_{L}=2c_{2}\rho^{2}${]}.
These notable relations are not well known. They deal with inhomogeneous
fluids and link linearly an excess free energy (times $\beta$) with
the corresponding excess adsorption. It is remarkable that Eq. (\ref{eq:inhomID})
does not include coefficients related to the interparticle potential.
Eq. (\ref{eq:inhomID}) resembles the equation of state of the bulk
ideal gas, nevertheless, it applies to any edge/wedge confined fluid
up to order $\rho^{2}$. As can be easily verified, the use of e-RR
provides more complex expressions for the surface and linear thermodynamic
properties than Eq. (\ref{eq:inhomID}). In summary, d-RR is appropriate
to obtain a simple description for the thermodynamics of the confined
ideal gas and also of any gas at low density, but e-RR is not.

\section{Application to Hard Spheres\label{sec:HSlowdens}}

Recently, through adopting the d-RR, the low density behavior of the
hard sphere (HS) confined fluid in an edge/wedge cavity was studied
using an analytic expression of $c_{2}(\alpha)$.\cite{Urrutia_2014b}
In this section we compare those properties with that found by adopting
the e-RR. With this purpose the natural units for the HS system will
be used (which is equivalent to set the particles diameter $\sigma$
as the unit length). In Ref. \cite{Urrutia_2014b} was obtained the
following exact expression 
\begin{equation}
c_{2}\left(\alpha\right)=-\frac{1}{15}\left[1+\left(\pi-\alpha\right)\cot\alpha\right]\:\label{eq:c2bet}
\end{equation}
that applies for $0<\alpha<\pi$ in the d-RR, while the analytic expression
for $\pi<\alpha<2\pi$ in d1-RR is 
\begin{equation}
c_{2}\!\left(\alpha\right)=\frac{8}{45}\left(\alpha-\pi\right)+Q\:,\label{eq:c2bet0}
\end{equation}
with $Q=0.007125\left\{ 1-\exp\left[-2.74\left(\alpha-\pi\right)\right]\right\} $.
Using the known parameters $b_{2}=-2\pi/3$, $a_{2}=-\pi/8$ and Eqs.
(\ref{eq:c2bet}, \ref{eq:c2bet0}) for $c_{2}\!\left(\alpha\right)$
one readily finds the series expansion of $\left\{ P,\gamma,\mathcal{T},\Gamma_{A},\Gamma_{L}\right\} $
in power of $\rho$ up to order two, by adopting both d-RR and d1-RR.
For $\mathcal{T}$ and $\Gamma_{L}$ it gives analytically the angular
dependence with $\alpha$ (up to order two in $\rho$).

Now, we analyze the consequences of choosing a different RR on the
thermodynamic properties of the confined HS system. In particular,
novel analytic expressions of relevant line-thermodynamic properties
for e-RR and d2-RR will be derived. Through the use of Eqs. (\ref{eq:intensY},
\ref{eq:adsY}), the matrices for RR transformation (\ref{eq:YmatMde},
\ref{eq:YmatMd1d2}, \ref{eq:YmatMd1e}), and the density power series
of $P_{\textrm{d}}$, $\gamma_{\textrm{d}}$, $\mathcal{T}_{\textrm{d}}$,
$\mathcal{T}_{\textrm{d1}}$, $\left(\Gamma_{A}\right)_{\textrm{d}}$,
$\left(\Gamma_{L}\right)_{\textrm{d}}$, $\left(\Gamma_{L}\right)_{\textrm{d1}}$
one obtains the series for $\gamma$, $\mathcal{T}$, $\Gamma_{A}$
and $\Gamma_{L}$ in the e-RR and d2-RR. For the wall-fluid surface
tension and excess area-adsorption, both up to terms of order $O(\rho^{3})$,
it is obtained
\begin{eqnarray}
\beta\gamma_{\textrm{e}} & = & \frac{\rho}{2}+\frac{5\pi}{24}\rho^{2}\:,\nonumber \\
\left(\Gamma_{A}\right)_{\textrm{e}} & = & -\frac{\rho}{2}+\frac{\pi}{4}\rho^{2}\:.\label{eq:HSsurfTense}
\end{eqnarray}
\begin{figure}
\begin{centering}
\includegraphics[width=0.95\columnwidth]{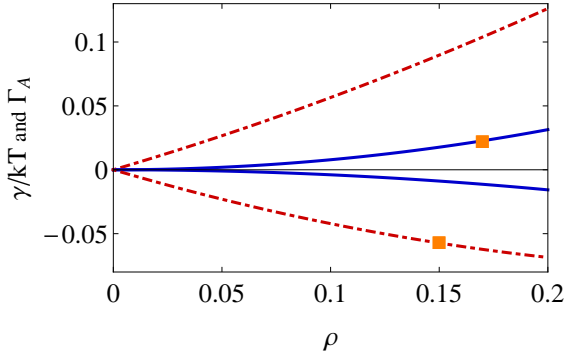}
\par\end{centering}

\protect\caption{Surface tension and surface adsorption. Curves for adsorption are
marked with squares. In continuous lines (blue) are plotted the magnitudes
in d-RR while dot-dashed lines (red) refers to e-RR.\label{fig:Surface-tension.}}
\end{figure}
Fig. \ref{fig:Surface-tension.} displays the surface tension and
surface excess area-adsorption by adopting d-RR and e-RR. There one
can observe the effect of choosing a different RR in the properties
of the confined HS fluid. For the d-RR one finds $\beta\gamma_{\textrm{d}}>0$
and $\left(\Gamma_{A}\right)_{\textrm{d}}<0$, on the other hand for
the e-RR they yield $\beta\gamma_{\textrm{e}}<0$ and $\left(\Gamma_{A}\right)_{\textrm{e}}>0$.
Besides, near $\rho\gtrsim0$ the null slope in $\beta\gamma_{\textrm{d}}$
and $\left(\Gamma_{A}\right)_{\textrm{d}}$ is apparent while $\beta\gamma_{\textrm{e}}$
and $\left(\Gamma_{A}\right)_{\textrm{e}}$ are linear with density.
For $\mathcal{T}_{\textrm{e}}$ in the range $0<\alpha<\pi$ there
are two branches: the first one for $\alpha<\frac{\pi}{2}$ and the
second one for $\alpha>\frac{\pi}{2}$, the corresponding expressions
are 
\begin{eqnarray}
\beta\mathcal{T}_{\textrm{e}} & = & -\frac{\rho}{4}\cot\!\frac{\alpha}{2}+\frac{\rho^{2}}{15}\left[1-\left(\frac{3\pi}{2}+\alpha\right)\cot\alpha-\frac{5\pi}{8}\csc\!\alpha\right]\nonumber \\
\beta\mathcal{T}_{\textrm{e}} & = & -\frac{\rho}{4}\cot\!\frac{\alpha}{2}+\frac{\rho^{2}}{15}\left[1+\left(\pi-\alpha\right)\cot\alpha-\frac{5\pi}{8}\cot\!\frac{\alpha}{2}\right]\:.\label{eq:HSlineTense}
\end{eqnarray}
\begin{figure}
\begin{centering}
\includegraphics[width=0.95\columnwidth]{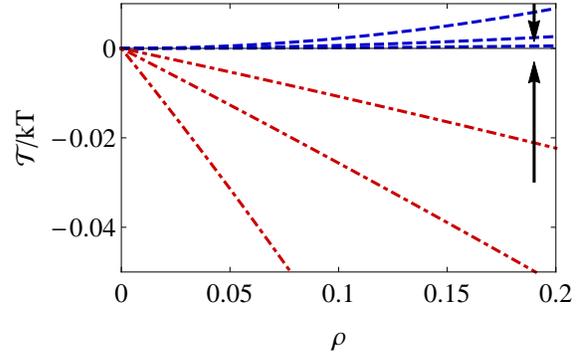}
\par\end{centering}

\protect\caption{Line tension of the HS system vs. density for angles in the range
$0<\alpha<\pi$. Continuous lines (blue) plot $\mathcal{T}_{\textrm{d}}$
while dot-dashed lines (red) plot $\mathcal{T}_{\textrm{e}}$. The
curves correspond to $\alpha=\pi/4,\,\pi/2,\,3\pi/4$ while the arrow
points to the direction of increasing values of $\alpha$.\label{fig:Lineten1}}
\end{figure}
In Fig. \ref{fig:Lineten1} it is shown the low density behavior of
the edge line tension $\mathcal{T}$ for d-RR and e-RR. In both cases,
$\mathcal{T}$ is monotonous. For the $\mathbf{M}_{\textrm{d}}$ measures
$\mathcal{T}$ is positive and has positive slope. On the contrary,
using $\mathbf{M}_{\textrm{e}}$ measures $\mathcal{T}$ is negative
and has a negative slope. For both, the modulus of the slope decreases
with increasing $\alpha$. The obtained expression of $\left(\Gamma_{L}\right)_{\textrm{e}}$
for $0<\alpha<\pi$ is
\begin{equation}
\left(\Gamma_{L}\right)_{\textrm{e}}=\frac{\rho}{4}\cot\!\frac{\alpha}{2}-\frac{2\rho^{2}}{15}\left[1+\left(\pi-\alpha\right)\cot\alpha+\frac{15\pi}{8}y(\alpha)\right]\:,\label{eq:HSlineAdse}
\end{equation}
which is non-derivable at $\alpha=\frac{\pi}{2}$ {[}see Eq. (\ref{eq:YmatMde}){]}.
\begin{figure}
\begin{centering}
\includegraphics[width=0.95\columnwidth]{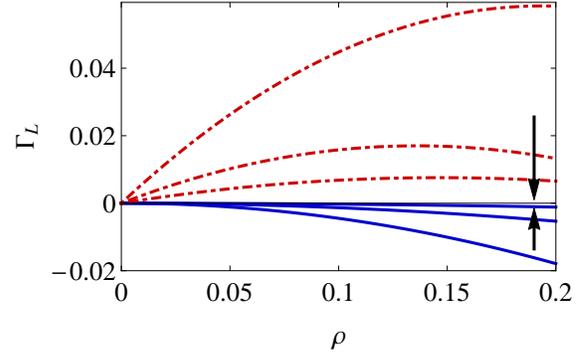}
\par\end{centering}

\protect\caption{Line adsorption of the HS system vs. density for angles in the range
$0<\alpha<\pi$, the curves correspond to $\alpha=\pi/4,\,\pi/2,\,3\pi/4$.
Continuous lines (blue) plot $\left(\Gamma_{L}\right)_{\textrm{d}}$
while dot-dashed lines (red) plot $\left(\Gamma_{L}\right)_{\textrm{e}}$.
See Fig. \ref{fig:Lineten1} for details.\label{fig:Lineads1}}
\end{figure}
Fig. \ref{fig:Lineads1} is similar to Fig. \ref{fig:Lineten1} but
for the linear adsorption $\Gamma_{L}$. When the $\mathbf{M}_{\textrm{d}}$
measures are considered the linear adsorption is monotonous, negative
(i.e., there is local desorption) and has negative slope which increases
with increasing $\alpha$. On the other hand, when $\mathbf{M}_{\textrm{e}}$
measures are adopted $\Gamma_{L}$ is not monotonous, it is positive
(i.e., there is local adsorption) and decreases for larger values
of $\alpha$.

I also present here a similar analysis for the case $\pi<\alpha<2\pi$.
The results for the line-tension and excess linear adsorption by adopting
e-RR and d2-RR up to order $O(\rho^{3})$ are
\begin{eqnarray}
\beta\mathcal{T}_{\textrm{e}} & = & \frac{\alpha-\pi}{8}\rho-\left[\left(\frac{\pi}{12}+\frac{8}{45}\right)\left(\alpha-\pi\right)+Q\right]\rho^{2}\:,\nonumber \\
\bigl(\Gamma_{L}\bigr)_{\textrm{e}} & = & -\frac{\alpha-\pi}{8}\rho+\left[\frac{16}{45}\left(\alpha-\pi\right)+Q\right]\rho^{2}\:,\label{eq:HSLineTensyAdse}
\end{eqnarray}
\begin{eqnarray}
\beta\mathcal{T}_{\textrm{d2}} & = & \left[\left(\frac{\pi}{8}-\frac{8}{45}\right)\left(\alpha-\pi\right)+Q\right]\rho^{2}\:,\nonumber \\
\bigl(\Gamma_{L}\bigr)_{\textrm{d2}} & = & \left[\left(-\frac{\pi}{4}+\frac{16}{45}\right)\left(\alpha-\pi\right)+Q\right]\rho^{2}\:,\label{eq:HSLineTensyAdsd2}
\end{eqnarray}
\begin{figure}
\begin{centering}
\includegraphics[width=0.95\columnwidth]{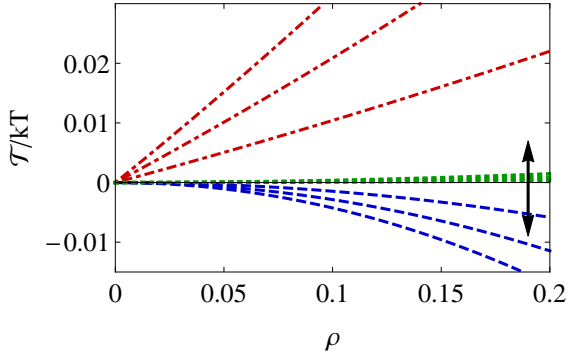}
\par\end{centering}

\protect\caption{Line tension of the HS system vs. density for angles in the range
$\pi<\alpha<2\pi$. The curves correspond to $\alpha=5\pi/4,\,3\pi/2,\,7\pi/4$.
In long-dashed lines (blue) it is drawn $\mathcal{T}_{\textrm{d1}}$,
short-dashed line (green) corresponds to $\mathcal{T}_{\textrm{d2}}$
and dot-dashed line (red) is for $\mathcal{T}_{\textrm{e}}$. The
arrows point to the direction of increasing values of $\alpha$.\label{fig:Lineten2}}
\end{figure}
Fig. \ref{fig:Lineten2} shows the linear tension of the HS system
in the case $\pi<\alpha<2\pi$. The functions $\mathcal{T}_{\textrm{d1}}$,
$\mathcal{T}_{\textrm{d2}}$ and $\mathcal{T}_{\textrm{e}}$ are shown
for comparison. In all cases, $\mathcal{T}$ is monotonous. For the
$\mathbf{M}_{\textrm{d}1}$ measures $\mathcal{T}$ is negative and
has negative slope which decreases with increasing $\alpha$. On the
other hand, for $\mathbf{M}_{\textrm{e}}$ and $\mathbf{M}_{\textrm{d}2}$
measures, $\mathcal{T}$ is positive and has a positive slope which
increases with increasing $\alpha$. Even, $\mathcal{T}$ for $\mathbf{M}_{\textrm{d}2}$
is nearly zero in the adopted scale. From the comparison between Fig.
\ref{fig:Lineten2} and Fig. \ref{fig:Lineten1} the inversion of the
sign of $\mathcal{T}_{\textrm{e}}$ at $\alpha=\pi$ is evident,
where the edge/wedge disappears.
\begin{figure}
\begin{centering}
\includegraphics[width=0.95\columnwidth]{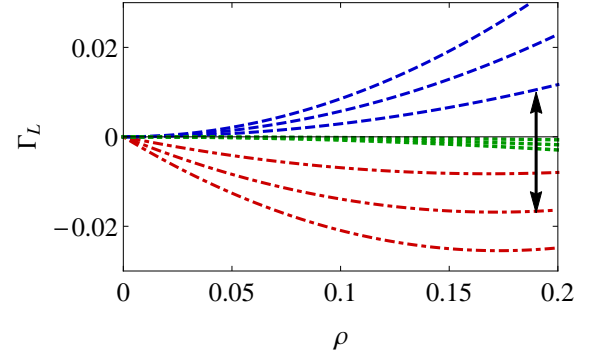}
\par\end{centering}

\protect\caption{Line adsorption of the HS system vs. density for angles in the range
$\pi<\alpha<2\pi$. In long-dashed lines (blue) it is draw $\left(\Gamma_{L}\right)_{\textrm{d1}}$,
short-dashed line (green) corresponds to $\left(\Gamma_{L}\right)_{\textrm{d2}}$
and dot-dashed line (red) is for $\left(\Gamma_{L}\right)_{\textrm{e}}$.
See Fig. \ref{fig:Lineten2} for more details.\label{fig:Lineads2}}
\end{figure}
 Fig. \ref{fig:Lineads2} plots the linear adsorption of the HS system
for the same three measure sets. In the cases of $\mathbf{M}_{\textrm{d}1}$
and $\mathbf{M}_{\textrm{d}2}$ we observe a monotonous $\Gamma_{L}$.
For the $\mathbf{M}_{\textrm{d}1}$ measures $\Gamma_{L}$ is a positive
(i.e., there is local adsorption) increasing function and its slope
increases with increasing $\alpha$. On the contrary, for $\mathbf{M}_{\textrm{d}2}$
measures $\Gamma_{L}$ is a negative (local desorption) decreasing
function and its slope decreases with increasing $\alpha$. Even,
$\Gamma_{L}$ for $\mathbf{M}_{\textrm{d}2}$ is nearly zero in the
adopted scale. For $\mathbf{M}_{\textrm{e}}$ measures $\Gamma_{L}$
is not monotonous, is negative (i.e., there is local desorption),
attains its minimum near $\rho\simeq0.17$ and decreases with increasing
values of $\alpha$. Fig. \ref{fig:Lineads2} and Fig. \ref{fig:Lineads1}
show the inversion of the sign of $\left(\Gamma_{L}\right)_{\textrm{e}}$
when both edge and wedge disappear at $\alpha=\pi$.

In the literature, both d-RR and e-RR were used to study HS systems
confined in cavities with different geometries.\cite{Roth_2010,Zhao_2012}
The behavior found with d-RR (including d1-RR and d2-RR) and e-RR,
for both line tension and line free energy, shows that they strongly
depend on the adopted reference system. Linear thermodynamic magnitudes
that are less dependent on the adopted reference region can be found
by considering the mean values of excess density and excess free energy
in a region with finite size around the edge.\cite{Urrutia_2014b}

\section{Final Remarks\label{sec:Conclusions}}

In this work I studied the relations between the thermodynamic properties
of fluids confined by wedges and edges when different RRs are adopted.
The analysis was based on the activity series expansion of the grand
free energy for inhomogenous systems, and on the properties of its coefficients
the cluster integrals. I utilized a simple approach that linearly
connects the geometric measures: volume, surface area and edge length,
in the different RR that are considered. From that, the law of transformation
of thermodynamic properties between RRs was deduced. A similar method
was previously used to study a system of hard spheres confined by
curved walls. The method was here refined and can be used to analyze
inhomogenous fluids confined by walls with a variety of shapes. Under
this non-standard approach I have studied the dependence of the linear-thermodynamic
properties on the adopted RR along the complete range of dihedral
angles $0<\alpha<2\pi$. Analytic expressions that transform the thermodynamic
intensive properties: pressure, surface tension and line tension of
the system when different RR are adopted were derived for the first
time. Surface adsorption and line adsorption were also analyzed in
this framework. The relevant results were given in Eqs. (\ref{eq:linetened},
\ref{eq:adsLed}, \ref{eq:linetend2d1}-\ref{eq:AdsLd2d1}, \ref{eq:linetened1}-\ref{eq:AdsLed1}).
Furthermore, the thermodynamic properties of both, the confined ideal
gas and of the confined real gases at low density (up to order $\rho^{2}$),
were analyzed by adopting different RR. We found that the density-based
d-RR is advantageous to obtain a simpler analytic description of the
studied properties. 

Regarding to the confined HS fluid, which is a relevant reference
system both for simple and colloidal fluids, the dependence of line
adsorption and line tension with the edge/wedge dihedral angle and
density was analyzed. We found explicit analytic expressions truncated
to order two in density that describe these properties for different
RR and for the complete range $0<\alpha<2\pi$. They are shown in
Eqs. (\ref{eq:HSlineTense}, \ref{eq:HSlineAdse}, \ref{eq:HSLineTensyAdse},
\ref{eq:HSLineTensyAdsd2}). The new results obtained for HS complement
those recently published.\cite{Urrutia_2014b} Given that these analytic
expressions are exact or quasi-exact, they constitute well defined
references that should enable to validate other approximate theories
like fundamental measure density functional approaches to edge/wedge
confined fluids at low density.
\begin{acknowledgments}
This work was supported by Argentina Grant ANPCyT PICT-2011-1887.
\end{acknowledgments}

\end{document}